# Ultra-compact tunable split-ring resonators

A. Vidiborskiy,[1] V. P. Koshelets,[1,2] L. V. Filippenko,[1,2] S. V. Shitov,[1,2] and A. V. Ustinov [1,3]

[1] National University of Science and Technology (MISIS), Leninsky prosp. 4, Moscow 119049, Russia

[2] Kotel'nikov Institute of Radio Engineering and Electronics (IREE RAS), Moscow 125009, Russia

[3] Physikalisches Institut, Karlsruhe Institute of Technology (KIT), 76131 Karlsruhe, Germany

**We propose tunable superconducting split-ring resonators (SRRs) employing nonlinear Josephson inductance. A fraction of SRR is replaced by Nb-AlO$_x$-Nb Josephson tunnel junctions connected in parallel and forming a superconducting quantum interference device (SQUID), whose inductance is sensitive to the external dc magnetic field. Due to the lumped nature of the Josephson inductance, the SRR can be made very compact and its resonance frequency can be tuned by applying magnetic field. We present the model, results of extensive EM-simulation and experimental data for the SRR weakly coupled to a transmission line within frequency range 11-13 GHz.**

Nowadays, metamaterials offer a wide range of possible applications[1]: cloaking devices, super lenses, antireflection coatings for solar cells, *etc*. One of the interesting areas includes metamaterials with negative index of refraction that requires simultaneously negative permittivity and negative permeability. To construct such a media, it is necessary to design special elements, which react with the magnetic and electric components of the incident wave. The split-ring resonators (SRRs) are often used for tailoring the magnetic reaction of the medium and, in general, have resonance frequency that is determined by dimensions of the rings. This kind of devices can be treated as meta-atoms.

Recent studies have been exploring different approaches to making split-ring resonators frequency-tunable[2-6]. The most conventional way is by using a semiconductor varactor as a tunable nonlinear capacitance[2-5]. The chip varactor has to be attached (e.g. soldered) directly to the planar structure. Such circuit can work in negative permeability regime near its resonance frequency, which can be shifted, for example, with infrared light[4].

Here, we propose superconducting SRRs with inductance enhanced by adding a chain of SQUIDs (Superconducting Quantum Interference Devices) formed by superconducting loops containing Josephson junctions (JJs). Our approach allows making such meta-atoms very compact and provides a possibility of tuning the resonance frequency by changing an external dc magnetic field. The inductance of the Josephson SRR (JSRR) is a periodic function of the magnetic flux threading the SQUID; the period is equal to magnetic flux quantum $\Phi_0 = h/2e \approx 2.07 \times 10^{-15}$ V·s, where $h$ is the Planck's constant and $e$ is the elementary charge. An idea of replacing a SRR in metamaterial by a SQUID having its intrinsic inductance and capacitance has been investigated earlier both theoretically[7-9] and experimentally[10]. In this paper, we explore an alternative approach, in which SQUIDs are used to increase solely the inductance of the conventional SRR and to make it tunable, leaving the characteristic resonator capacitance unaffected. Hereby, the frequency of the intrinsic SQUID resonance remains well above the SRR resonance frequency.

To facilitate the design, we started with analysis of a JSRR approximated as a lumped *LC*-circuit[11] shown in Fig. 1(a). The geometrical inductance $L_g$ is defined by inductance per unit of length of two parallel strips and by their mutual inductance. The geometrical capacitance of a SRR, $C_g$, is given by the sum of two components, the capacitance of the gap and the capacitance between two strips separated by a dielectric. To measure the resonance frequency, the JSRR was placed aside a 50-Ohm microstrip transmission line. The transmission line, which plays the role of a readout circuit, is made narrower within the coupling region. The microstrip introduces extra capacitance of about 7% of the total JSRR capacitance, which produces almost negligible change of the resonance frequency. The structure, including the microstrip, is fabricated as Nb thin film on a silicon substrate. Josephson junctions are produced using the conventional Nb-AlO$_x$-Nb trilayer process. The silicon chip is attached to a printed circuit board (PCB) with conducting glue on top of the ground plane made of copper. The PCB provides connection between coaxial cables from/to a vector network analyzer and the micro-strip line on the chip. The sample holder is placed inside a magnetic coil to provide dc magnetic field to the SQUIDs. The whole assembly is placed inside a µ-metal shield (cryoperm) and installed at 1.2-K stage of a dry cryostat.

The JSRR has been simulated using the AWR MWO environment for the layout parameters of experimental structure. Since SRR acts effectively as a semi-lumped resonator, the anti-node of the standing wave is located in the center of the outer ring. Thus, the largest effect of adding an extra inductor can be expected when placing it in the middle of the outer part of the SRR.

Since the standard MWO EM-simulator does not support direct implementation of the Josephson inductance, we use the electrical equivalent of JJ depicted in Fig. 1(a). This linear model of a JJ should be valid for small probe signals. The SQUID can be represented[12] as a parallel connection of a capacitor, which substitutes the capacitance of the tunnel barrier, and a magnetic-field dependent Josephson inductor. The Nb-AlO$_x$-Nb Josephson junctions are fabricated with the following parameters[13]: junction area $S$=7 μm$^2$, specific tunnel resistance $R_{na}$=1500 Ohm·μm$^2$ and energy gap voltage $\Delta$=2.8 mV. The maximum critical current of the junction is estimated using the Ambegaokar-Baratoff relation[14]:

$$I_c^{A.-B.} = \frac{\pi\Delta}{4R_n}, \tag{1}$$

where $I_c^{A.-B.}$ is the critical current, $R_n$ is the normal state resistance of the junction with an area of $S$, and $\Delta \cdot e$ is the energy gap of superconducting Nb. The maximum superconducting current of the junction is estimated as $I_c^{A.-B.} = 10.25$ μA. The variable Josephson inductance is given by expression[15]:

$$L_J = \frac{\Phi_0}{2\pi I_c \cos\varphi}, \tag{2}$$

where $\varphi$ is the superconducting phase difference across the junction. In our case, each SQUID consists of two JJs connected in parallel. The following parameters are estimated for each of the used Josephson junctions: $L_J(0)$=32 pH, $C_J$=490 fF. Thereby, the plasma frequency for the junction $f_p=1/2\pi(L_J C_J)^{-1/2}\approx$40 GHz which characterizes the resonant frequency set by the junctions' own capacitance and Josephson inductance, is well above the operating frequency of our circuit. Thus, the intrinsic resonance of JJ is not expected to influence the following measurements.

We have chosen the operating frequency range of SRR near 11.5 GHz. The EM-simulation demonstrated that embedding a series array of SQUIDs into a SRR makes it possible to significantly decrease the dimensions of the split ring for a frequency of choice. Fig. 1(b) shows the comparison between conventional SRR and JSRR, both designed for the same operation frequency. Adding 15 SQUIDs into a SRR reduces its footprint by a factor of about 3. Moreover, the area of metallization is reduced 6 times, which is favorable for minimizing the metal density of an artificial electromagnetic media.

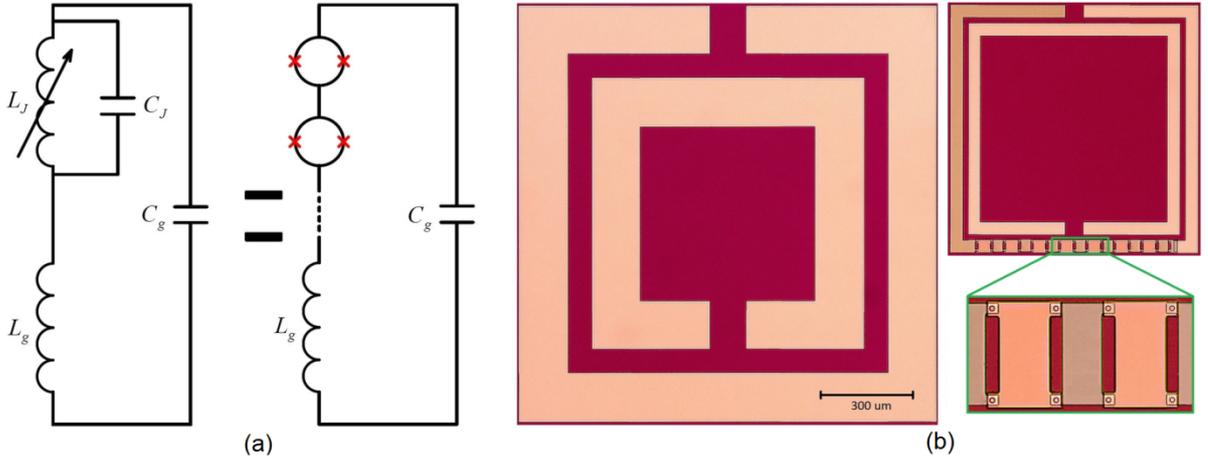

FIG. 1. (a) Equivalent circuit for the JSRR, which integrates a few SQUIDs. Here $L_g$, $C_g$ are the geometrical inductance and capacitance of the JSRR respectively. The (red) crosses indicate Josephson junctions with inductance $L_j$ and capacitance $C_j$. (b) Size comparison of the fabricated conventional SRR and tunable JSRR containing 15 SQUIDs. Both meta-atoms where designed to operate around 11.5 GHz and are shown using the same scale. Bottom inset shows a 5-time enlarged fragment of the SQUID array; small circles are Josephson junctions.

The results of measurement for the transmission coefficient $S_{21}$ versus the applied dc magnetic field are shown with a color scale in Fig. 2(a, b). The resonance frequency varies periodically as a function of magnetic flux applied to the sample. The frequency maxima are at the $n\Phi_0$ flux and minima at $(n+1/2)\Phi_0$, where $n$ is an integer. These periodical variations originate from $2\pi$ periodicity of Eq.(2) in $\varphi$. The minima of the resonance frequency are limited by the screening parameter of the SQUID[16] $\beta_L=2L_gI_c/\Phi_0$. This parameter restricts the variation of the phase difference $\varphi$ in the denominator of Eq. (2). The accurate value of $L_g$ has been extracted from the detailed EM simulations of our structure. For the samples with 9 and 15 SQUIDs shown in Fig. 2(a) and (b), the inductance is $L_g^9 \approx 60$ pH, $L_g^{15} \approx 28$ pH respectively. In this case, we obtain $\beta_L^9 \approx 0.54, \beta_L^{15} \approx 0.25$, meaning that the critical current of each SQUID can be varied by not more than factor of 2.5 and 4, respectively. The value of the critical current obtained from the numerical fit to the experiment is $I_c$=9.4 μA, which is 8% less than the value estimated from the fabrication parameters according to Eq.(1), that can be explained by the strong coupling effects in niobium.

From Fig. 2 one can see that the depth of modulation is being reduced in stronger fields for both positive and negative directions of the dc magnetic field. Such behavior can be explained by non-uniform spread of local magnetic fluxes over the SQUIDs. This may occur due to the Meissner effect in superconducting electrodes and thus a difference between demagnetization factors of SQUIDs. Consequently, the Josephson inductance of the SQUIDs does not change equally in the common magnetic field. This is especially important at $(n+1/2)\Phi_0$ flux, when the denominator in Eq. (2) dramatically drops. The effect becomes more pronounced in larger magnetic field, as it is clearly seen in Fig. 2 (b) in the region of ±10 μT. The best fit to the experimental data shown in Fig. 2 (black dashed lines) is obtained via introducing a demagnetization factor profile according to the empirical formula:

$$L_n(\Phi) = \frac{L(0)}{\left|\cos[\pi \cdot \Phi / \Phi_0 \cdot (1-nC)]\right|}, \qquad (3)$$

where $n$ is an integer number defining the position of the SQUID in respect to the axis of symmetry, $C$ – empirical coefficient, and $\Phi$ – the total flux threading the SQUID. The best fit to experimental data yields $C$=0.0135 as shown in Fig. 2.

The fitting of the effective inductance described above gives us the value for the minimum of Josephson current of about 4 μA. The probing microwave signal in our experiment was set to −80 *dBm*. According to the results of our EM-modeling, this power level corresponds to the ac current of 3.2 μA through the JJ at the resonance that is comparatively large to cause nonlinear effects in the SQUIDs. The above estimate for the minimum critical current of the experimental SQUIDs at $(n+1/2)\Phi_0$ flux is rather close to this value. Thus, nonlinearity of SQUIDs can be yet another reason for the minimum frequency of JSRR measurable in our particular experimental conditions. As an extra illustration, dotted lines in Fig. 2 represent the fit obtained from EM-simulation in the limit of extremely small probe signal and disregarding non-uniformity of the flux given by Eq.(3).

It can be seen from Eq. (2) that the Josephson inductance has its minimum at either zero magnetic field or at integer $\Phi_0$ flux values. The observed offset of the minima with respect to zero field (e.g. about $0.4\Phi_0$ for Fig. 2(b)) is due to magnetic flux trapped in thin niobium films[17]. This flux in the form of Abrikosov vortices is concentrated at inhomogeneities[18] of the films such as vias, fringes, steps of superconducting films and their interconnections.

In conclusion, we reported on new type of the tunable split-ring resonators. Experimental results are in good quantitative agreement with our analysis and EM-simulations. Transforming traditional SRR into JSRR offers advantages of much lower losses, frequency tunability, significantly reduced area of metallization and footprint of the meta-atom. This all makes the JSRRs promising substitute of the conventional SRRs, in particular for applications at cryogenic temperatures.

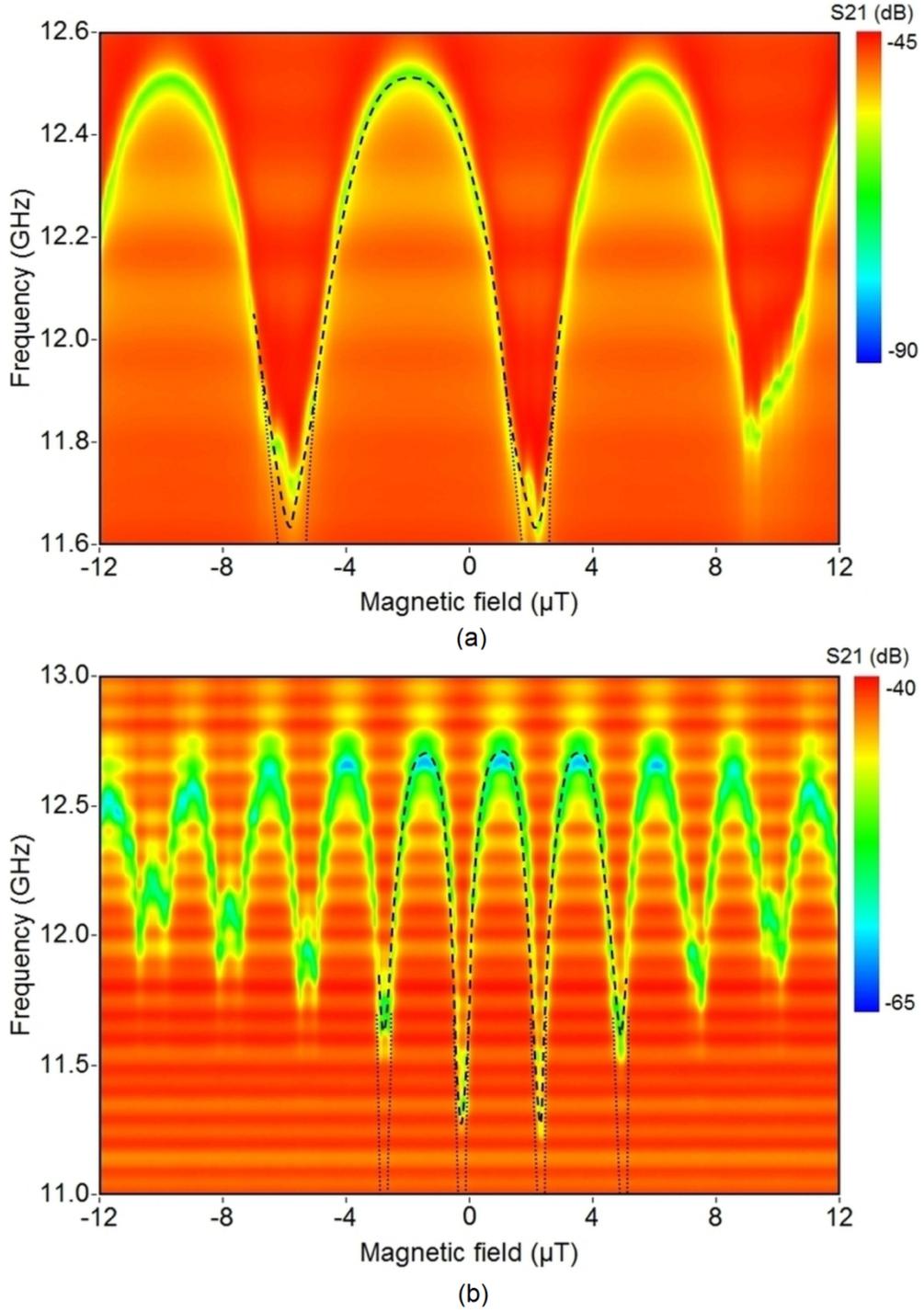

FIG. 2. Transmission coefficient $S_{21}$ measured across the readout microstrip line. (a) JSRR with 9 SQUIDs, $P_{signal} \approx -75$ dBm, $\beta_L=0.54$ (b) JSRR with 15 SQUIDs, $P_{signal} \approx -80$ dBm, $\beta_L=0.25$. The black dashed lines are fits for non-uniform field according to Eq.(3) and saturation limit of the critical current of 3.2 µA. The dotted lines represent the fit for the negligibly small probe signal and uniform magnetic field.

Authors would like to acknowledge stimulating discussions with S. M. Anlage, A. S. Averkin, P. Jung, G. Tsironis, E. Zharikova and thank N. Abramov for technical assistance. This work was supported in part by the Ministry of Education and Science of the Russian Federation and the Russian Foundation of Basic Research.